# The damped oscillating propagation of the compensating self-accelerating beams

Wei-Wei Liu[1], Yong-Lun Jiang[1], Pan-Pan Yu[1], Hao-wei Wang[1],

Zi-qiang Wang[1] and Yin-Mei Li [1,*]

[1] Department of Optics and Optical Engineering, University of Science and Technology of China, Hefei, 230026, P.R. China
*Corresponding author: liyinmei@ustc.edu.cn

We report a new form of compensating accelerating beam generated by amplitude modulation of the symmetric Airy beam (SAB) caustics with an exponential apodization mask. Our numerical study manifests that the compensating beam is with one main-lobe beam structure and can maintain the mean-intensity invariant both in the free space and loss media. Specially, the beam inherits the beamlets structure from the SAB and owns a novel damped oscillating propagation property. We also conduct a comparative study of its propagation property with that of the Airy beam theoretically. And by altering the signs of 2D masks, the main lobe of the compensating beam can be modulated to orientate in four different quadrants flexibly. We anticipate the proposed compensating accelerating beam will get special applications in particle manipulation or plasmas etc. region.

## Ⅰ. INTRODUCTION

The self-accelerating beams have attracted intense interest in theoretical and applied researches since the Airy beam was theoretically and experimentally demonstrated [1-3]. As paraxial solutions to the Schrodinger equation discovered by Berry and Balazs[3], the Airy beam with parabolic travelling trajectory broken the inherent impression that the light propagates in straight lines in free space. Other types of self-accelerating beams with different patterns or propagation paths were also proposed for instance, non-paraxial elliptical propagation Mathieu and Weber beams [4], three dimensional accelerating beams [5] and accelerating beams with arbitrarily transverse shapes [6]. Distinguished from the commonly straight light, these accelerating beams all travel in curved trajectories and have been exploited in large amounts of intriguing applications such as optical clearing of micro particles [7], light-sheet microscopy [8, 9], plasma physics [10] and even generating self-bending electron beams [11] or light bullets [12]. Specially, by shaping the amplitude and phase of light field, the accelerating beam can even be tailored to travel along arbitrary trajectories [13-16]. And they [all of these self-accelerating beams] are endowed with non-diffraction and self-healing properties which allow the beams to preserve the invariant intensity profile during the bending propagation and can reconstruct themselves after passing through partially blocking or distortion. [17]

It's known that ideal accelerating beams were defined to convey infinite energy and were claimed to own diffracting-free property. Actually, the practical accelerating beams generated from truncated aperture experimentally are with finite energy and can only maintain the non-diffracting properties in limited distance [1, 2, 4, 18]. Especially, when the finite-energy accelerating beams propagate in absorbing media, the absorption will lessen the beams propagation distance and may even influence their optical functionality in many applications such as the generation of Airy Plasmon [10, 19] or optical acceleration for microparticles [17]. For a long time, the solution for the issue of absorption was only through amplifiers or other external means but its function was limited to extend the beam propagation distance and hard to make the beams' structure keep invariant [17]. However, recent exploration showed that adjusting the spectral properties of self-accelerating beams was an effective way to obtain the shape-preserving accelerating beams in loss media [17, 20].

In this paper we propose one kind of apodization mask for amplitude modulation of symmetric cubic phase to obtain a new form of compensating self-accelerating beams (compB) numerically. The mask aims to adjust the caustics distribution of symmetric airy beam (SAB) in spectral plane and is capable of transferring the energy from oscillating tail of SAB to concentrate on its one off-axis lobe, thus creating the beam with just one-side main-lobe as that of the Airy beam. Differently, the compensating accelerating beam can propagate in both free space and absorbing media maintaining the beam structure and mean intensity invariant in an extended distance. Specially, it also inherits the beam structure from the SAB and exhibits the novel damped oscillation propagation property for its main-lobe along the bending trajectory. Compared with the previous smooth [17] or periodic [21] accelerating beams, it's the first time to show the self-accelerating beams own the damped oscillating propagation property. At last, we compare the propagation profiles of this accelerating beam with that of Airy beam in 1D and 2D cases for further illustrating its propagation features. We anticipate that the created compensating accelerating beams could benefit related applications such as optical accelerating, optical micromanipulation and fluorescence imaging regions.

## II. THEORETICAL FRAMEWORK

We begin with demonstrating the theoretical analysis of the proposed compensating self-accelerating beam generated from SAB. Based on angular spectrum formalism [22], the paraxial solution $u_0(x,z)$ for finite-energy 1D SAB propagating in the $z$ direction can be expressed as:

$$u_0(x,z) = C\int_{-\infty}^{+\infty} dK U_0(K) e^{-aK^2 x_0^2} e^{iz\sqrt{k^2-K^2}} e^{ixK} , \quad (1)$$

where $C = x_0 e^{(a^3/3)}$ is a constant and $U_0(K)e^{-aK^2 x_0^2} = F\{u_0(x,0)\}$ is Fourier transformation of $u_0(x,z)$ at $z=0$ with conjugate coordinate $K$, the parameter $x_0$ refers to the beam transverse size as in the case of Airy beam [1, 2]. $k = 2\pi/\lambda$ is the wave number in vacuum. Specifically, $e^{-aK^2 x_0^2}$ is Gaussian amplitude with parameter $a>0$ to guarantee the square integrality of Eq. (1), and $U_0(K)$ is the angular spectrum "symmetric cubic phase" which generates SAB. It's with even parity and reads

$$U_0(K) = e^{\frac{i|K|^3 x_0^3}{3}} . \quad (2)$$

Altering the angular spectrum properties of self-accelerating beams can adjust their caustics distribution and eventually influence the beams' propagation property [17, 20, 23]. Here the new compensating self-accelerating beam is created through pupil apodization on the angular spectrum plane of SAB with exponential amplitude mask $\exp(-bK)$, where $b$ is a variable real parameter. The mask can make the spectral components of symmetric cubic phase distribute exponentially along the transverse direction and thus impel the beam's caustics to be more concentrated in the one main-lobe of SAB to realize the compensating effect. Imposing the mask in angular spectrum plane of SAB and after propagating a distance z in linearly absorbing media, the beam spatial spectrum becomes:

$$U(K,z) = U_0 \exp(bK)\exp(iz\sqrt{k^2-K^2} - \alpha z), \quad (3)$$

where $\alpha$ is the absorption coefficient of the medium. It should be emphasized that the modulation of this apodization mask is based on finite-energy condition so the Gaussian amplitude in angular spectrum plane is still needed. Therefore based on angular spectrum formulism similarly the paraxial solution for the above compensating self-accelerating beam can be expressed:

$$u(x,z) = C\int_{-\infty}^{+\infty} dK U(K,z) e^{-aK^2 x_0^2} e^{ixK} . \quad (4)$$

The compensating accelerating beam inherits the beam structure from SAB but with just one-side main lobe as the Airy beam. That is, the compensating beam also exhibits the propagation properties of damped oscillating in both free space and in loss medium. What's more, the beam's mean intensity of main lobe can maintain propagation-invariant in various absorbing media by setting an appropriate value of the parameter b. In the following text we will illustrate them based on the numerical results from Eq. (4).

## III. DYNAMIC PROPAGATION PROPERTIES OF THE COMPENSATING ACCELERATING BEAMS

### A. In the free space

Based on the numerical analysis from Eq. (1) and Eq. (4), a comparison between SAB and the compensating accelerating beam with $b=48\mu m$ which maintains the mean-intensity for its main-lobe, both propagating in free space, is depicted in Fig. 1. Fig.1 (a) shows the intensity profile for 1D SAB as a function of $(s,\xi)$, and its transverse intensity distribution at the typical planes $\xi = 2.5, 6$ is depicted in fig.1(c), here $s$ and $\xi$ are dimensionless transverse and longitudinal parameters: $s = x/x_0$, $\xi = z/kx_0^2$. It can be seen that except the central lobe with autofocusing character, the SAB also owns two off-axis lobes which shape like two conventional Airy beams but with discrete beamlets along the parabolic trajectories. However, the intensity of its two axis-lobes decays quickly during propagation which is illustrated by the substantial intensity attenuation in fig.1(c) due to the finite-energy condition [22].

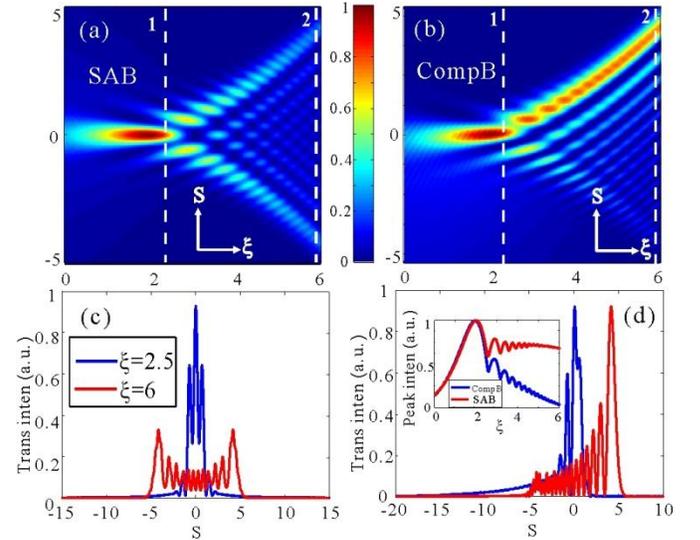

Fig.1. The intensity distribution for 1D (a) SAB and (b) the compensating accelerating beam with $b=48\mu m$ propagating in free space as a function of $(s,\xi)$ numerically, $s \in [-5,5]$ and $\xi \in [0,6]$. The intensity scale is normalized to the local maximum intensity for each snapshot. The transverse intensity distribution for above (c) 1D SAB and (d) the compensating accelerating beam at their typical planes $\xi = 2.5, 6$ as a function of transverse coordinate $s$. Inset in fig.(d): the peak intensity for above normalized 1D SAB and compensating accelerating beam as a function of $\xi$. The beam parameters for above 1D beams are all the same with $\lambda = 532nm$, $x_0 = 200\mu m$ and $a = 0.05$.

On the other hand, the paraxial compensating accelerating beam is with one main lobe beam structure which resembles the Airy beam [1] but maintains the intensity and shape along the bending trajectory (fig. 1(b)). Comparatively, fig. 1(d) illustrates its propagation-invariant property by showing the peak intensity of its main-lobe at the typical planes $\xi = 2.5, 6$ being identical. This is because the increased caustics from the oscillating tail between $S \in [-5, 0]$ (fig.1(c)) are transferred to the compensated main lobe modulated by the exponential mask, accompanying the other lobe to disappear. For another, the modulated beam also inherits the structure from SAB with the central lobe at the initial propagation and the beamlets structure along the main lobe. Specially, the compensating beam possesses the novel damped oscillating propagation property as a consequence of the discrete intensity distribution of the beamlets, which has not been reported before. The inset in fig. 1(d) shows the peak intensity of the above 1D SAB and compensating accelerating beam in free space. Obviously, their maximum intensity both occurs in the central lobe. After that, the SAB decays exponentially in intensity, while the compensating beam can maintain its mean intensity with damped oscillating propagation property until the propagation plane $\xi = 6$.

### B. In the absorbing media

As the exponential mask can modulate the beam propagation in the free space efficiently, it can be inferred that enhancing the compensation by increasing the value of the mask parameter $b$ could make up to the extra loss to obtain the propagation-invariant accelerating beam in absorbing media. And fig. 2(a) combining with fig. 2(c) show the targeted accelerating beam in the linear absorbing media by setting the mask parameter $b = 70\mu m$ and the absorption coefficient $\alpha = 0.117 dB/m$. It can be seen that the compensating beam in loss media also inherits the beam structure from the SAB with damped oscillating propagation property as it does in free space (fig. 1(b)). Moreover, fig.3(c) shows that it also has the same peak intensity at the planes $\xi = 2.5, 6$, which indicates its propagation property. On the other hand, note that the boundary between the beamlets in the beam main-lobe becomes vague since more caustics concentrate on the lobe modulated by the exponential mask.

Fig.2. The intensity distribution for 1D (a) propagation-invariant compensating accelerating beam with $b = 70\mu m$ exists in loss media with absorption coefficient $\alpha = 0.117 dB/m$ and (b) oscillating-growing beam with $b = 70\mu m$ in lossless media as a function of $(s, \xi)$ numerically, $s \in [-9, 9]$ and $\xi \in [0, 8]$. The intensity scale is normalized to local maximum intensity for each snapshot. Transverse intensity distribution for above (c) 1D propagation-invariant compensating accelerating beam and (d) oscillating-growing beam at the typical planes $\xi = 2.5, 6$ as a function of transverse coordinate $S$. Inset in (d): normalized peak intensity for above 1D beams as a function of $\xi$. The beam parameters for the 1D beams are all the same in Fig.1.

An interesting character of the compensating accelerating beam with $b = 70\mu m$ is that it shows oscillating-growing property when propagating in lossless media ($\alpha = 0$), which the beam is still shape-preserving with oscillating propagation property but the intensity of its main-lobe is increasing along the curve (fig. 2(b)). The result is an over-compensating effect, which results from that the power transformation from the caustics between $S \in [-5, 0]$ to the main-lobe is excessive truncated by the apodization mask. The inset of fig. 2(d) shows the peak intensity of above 1D compensating beam in loss and lossless media separately. Evidently, both of them display oscillating property during the propagation and illustrate the propagation dynamics of the compensating beam with $b = 70\mu m$ in different media. It can be seen that after the intensity fluctuation of the central lobe, the beam can keep intensity oscillating-rising until the plane $\xi = 7$. Comparatively, it has the same propagation behavior as the compensating beam with $b = 48\mu m$ does in the free space. The special oscillating structure of the compensating accelerating beam may get special application such as in particle accelerating where we can simulate the oscillating behavior and therein the central lobe could even act torque-amplification function for the particles. The oscillating-growing beam is also useful in some applications which need the highest intensity of the beam main lobe to reach enough far targeted position.

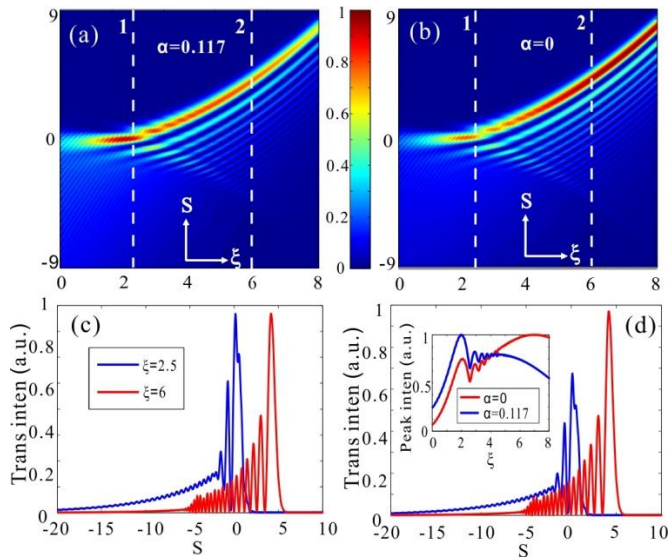

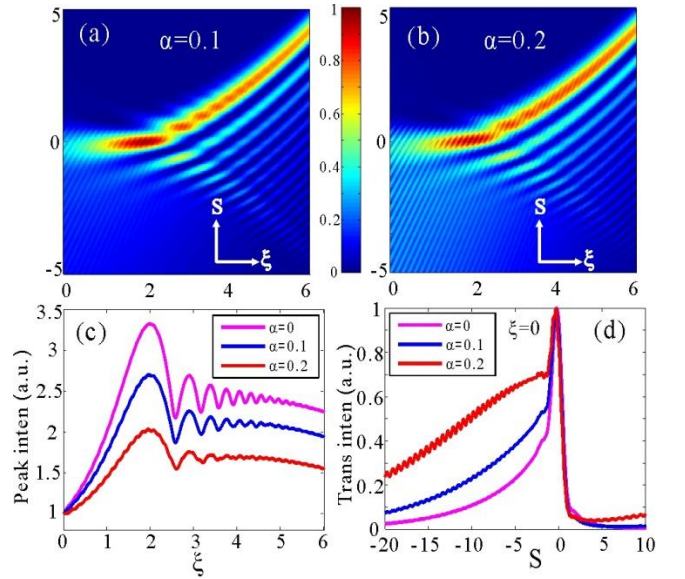

Fig.3. The intensity distribution for the 1D propagation-invariant accelerating beams in loss media with parameter (a) $b = 63\mu m$, $\alpha = 0.1 dB/m$ and (b) $b = 84\mu m$, $\alpha = 0.2 dB/m$ as a function of $(s, \xi)$ numerically, $s \in [-5,5]$ and $\xi \in [0,6]$. The intensity scale is normalized to the local maximum intensity for each snapshot. Transverse intensity distribution for above (c) The normalized peak intensity to the peak intensity at initial plane $\xi = 0$ for the 1D propagation-invariant accelerating beams in the media with absorption coefficient $\alpha = 0, 0.1, 0.2 dB/m$ as a function of $\xi$. (d) The normalized transverse intensity for the above 1D beams at plane $\xi = 0$ as a function of transverse coordinate $S$. The beam parameters for the beams are all the same in Fig.1.

Actually, through setting appropriate compensating value of the parameter b, we can always obtain the propagation-invariant oscillating beam in arbitrary absorbing media. Here we compare the propagation behavior of these beams in different media. And fig. 3(a) and 3(b) show the propagation dynamics of the additional 1D propagation-invariant beam in loss media with the absorption coefficient $\alpha = 0.1 dB/m$ and $\alpha = 0.2 dB/m$ separately, the mask compensating parameters for the beams are $b = 63\mu m, 84\mu m$ correspondingly. Firstly, it can be seen that they also inherit the beam structure from the SAB with propagation-invariant and damped oscillating property. Furthermore, note that the boundary between the beamlets along the main-lobe in fig. 3(b) is vaguer than that in fig. 3(a) as more caustics are modulated to concentrate on the main-lobe to make up to the extra loss. For comparing their propagation dynamics, fig. 3(c) shows the normalized peak intensity of the above two beams as well as the propagation-invariant beam in the lossless media. It's identified that the oscillating amplitude of the beams' central lobe and beamlets in lossless media is the largest, while as the medium absorption increases, their oscillating amplitude especially the central lobe is attenuated correspondingly (fig.3(c)). On the other hand, fig. 3(d) gives the normalized transverse intensity of the above beams at the initial plane $\xi = 0$, the beams intensity during $S \in [-20,0]$ is apparently proportional to the medium absorptivity, which indicates that the larger absorption of the medium, the more caustics are concentrated on the beam's one side modulated by the mask to keep the beam propagation-invariant.

### IV. COMPARING THE PROPAGATION CHARACTER OF THE COMPENSATING ACCELERATING BEAM AND THE AIRY BEAM

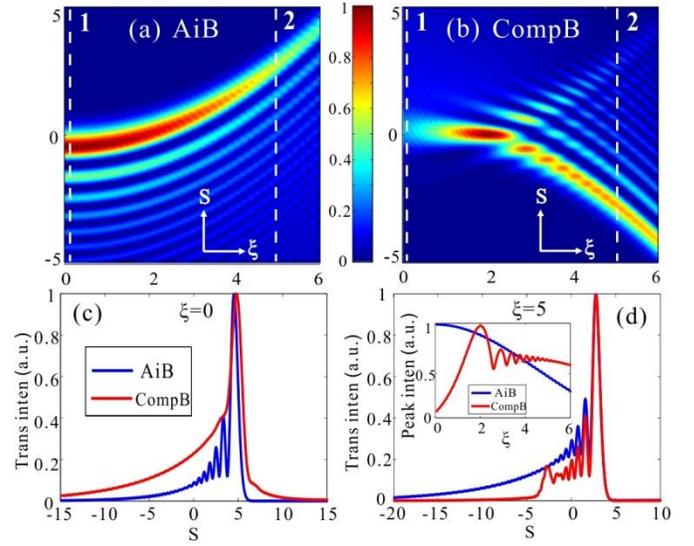

Fig.4. The intensity distribution for the 1D (a) Airy beam and (b) propagation-invariant accelerating beam exists in free space with $b = -48\mu m$ and a function of $(s, \xi)$ numerically, $s \in [-5,5]$ and $\xi \in [0,6]$. The intensity scale is normalized to local maximum intensity for each snapshot. The normalized transverse intensity distribution for above 1D Airy beam and propagation-invariant accelerating beam with $b = 48\mu m$ at propagation plane (c) $\xi = 0$ and (d) $\xi = 5$ as a function of transverse coordinate $S$. Inset in (d): Peak intensity for above normalized 1D Airy beam and propagation-invariant accelerating beam as a function of $\xi$. The beam parameters for above 1D beams are all the same in Fig.1.

For further revealing the feature of the compensating accelerating beam, we compare its propagation dynamics with that of the Airy beam [1]. Fig. 4(a) and 4(b) show the intensity patterns of the 1D finite-energy Airy beam and the propagation-invariant accelerating beam with the negative mask parameter $b = -48\mu m$ in the free space as a function of $(s, \xi)$, the two beams share the same beam parameters $a, x_0$ and $\lambda$ as that in fig.1. Firstly, the compensating accelerating beam obviously owns the same parabolic trajectory as the Airy beam's. Distinguishingly, the energy distribution of the Airy beam is smooth along the bending trajectory and the intensity of its main-lobe decays on propagation (fig. 4(a)), while the compensating beam can maintain its mean-intensity and owns the beamlets structure accompanied by the damped oscillating property (fig. 4(b)). Note that the energy distribution of the compensating beam in fig. 4(b) is exactly the same as that in fig. 1(b) along their parabolic trajectory but distributes in the other side of the beam profile transversely about the horizontal axis $s = 0$, which results from the caustics concentration in $s = [0,5]$ modulated by the negative mask. This indicates that the orientation of the compensating beam main lobe can be altered by the mask and we will illustrate it in detail in the following 2D case. For clearly contrasting their caustics distribution, fig. 4(c) and 4(d) display the normalized transverse intensity for the 1D Airy beams and the propagation-invariant beam with $b = 48\mu m$ as a function of $s$ at the planes $\xi = 0, 5$. One can figure out that the caustics of the propagation-invariant beam distribute similarly in shape with that of the Airy beam. However, its intensity (in red) at the initial plane $\xi=0$ is higher than that of the Airy beam (in blue) in fig. 4(c) but lower at the plane $\xi=5$ shown in fig. 4(d), which indicates that the caustics contribution to the main lobe and the main-lobe energy concentration of the compensating beam are both higher than the situations of the Airy beam. The inset in fig. 4(d) shows the normalized peak intensity of the two beams during propagation as a function of $\xi$. Obviously, comparing with the decaying tendency of the Airy beam, the result for the propagation-invariant accelerating beam confirms the compensating function of our proposed apodization mask.

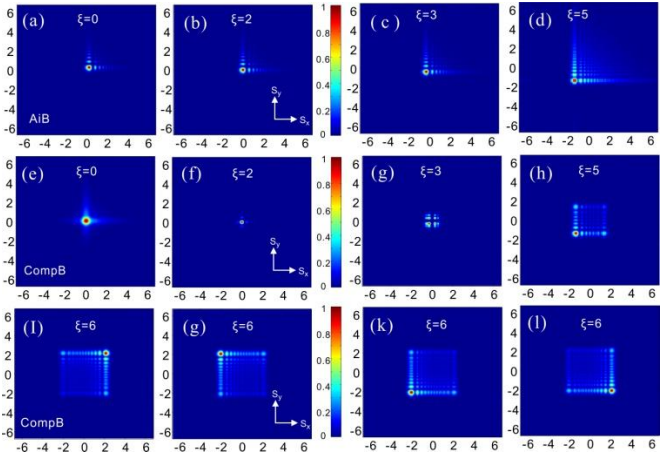

Fig.5. Theoretical intensity patterns for the 2D Airy beam (first row) and compensating accelerating beam with $b=48\mu m$ (second row) at their respective planes $\xi=0,2,3,5$. The above intensity profiles (third row) of compensating accelerating beam with their main lobes in four different quadrants modulated by the apodization masks $\exp(\pm bK_x \pm bK_y)$ with different signs. The intensity scale is normalized to local maximum intensity for each snapshot. The beam parameters for the above 2D beams $\lambda=532nm$, $x_0=200\mu m$ and $a=0.08$.

Fig.5 shows the 2D transverse profiles of the Airy beam (first row) and the compensating accelerating beam with $b=48\mu m$ (second row) in the free space numerically as a function of $(S_x,S_y)$ at the planes $\xi=0,2,3,5$. It can be seen that the central lobe of the 2D compensating beam also autofocus around the plane $\xi=2$ as the SAB does [22], but for another, only one apex lobe of the 2D beam profile in the third quadrant owns the maximum intensity which is similar with the energy distribution of the 2D Airy beam. Meanwhile, after the autofocus, the survived apex lobe evolves along the transverse direction as the way the Airy beam does. Note that the transverse profile width of the 2D compensating beam is larger than the Airy beam's at the initial plane $\xi=0$ but narrower at the other planes $\xi=2,3,5$. This is due to the compensating effect caused by the 2D apodization mask that more caustics distribute at the initial plane and the energy concentration for the main-lobe of the compensating beam is also higher during propagation as already illustrated in fig.4(c) and 4(d). Furthermore, as the beam energy flow direction can be adjusted by altering the sign of the apodization mask (fig. 4(b)), here the 2D masks are set with different signs identically to adjust the orientation of the beam apex lobe. The third row of fig.5 shows the intensity profiles of the compensating beams at the plane $\xi=6$ modulated by four kinds of 2D masks with the expression being $\exp(\pm bK_x \pm bK_y)$ respectively, where $b=48\mu m$, and one can see that the apex lobe of the compensating beam can be modulated to orientate in four different quadrants.

## V. CONCLUSION

In conclusion, we have proposed a new form of compensating accelerating beam by amplitude modulation of the SAB caustics in the angular spectrum plane with an exponential apodization mask. The compensating beam is with one-side main lobe beam structure, and it can not only maintain the beam mean-intensity invariant during the parabolic propagation but also inherits the center lobe and beamlets structure from the SAB with the damped oscillating property. Our numerical results showed that the beam can maintain the propagation-invariant and damped oscillating propagation property in different absorbing media. And also the damped oscillating amplitude of the propagation-invariant beam in loss media is inverse proportion to the medium absorptivity when the compensating beam get the same peak intensity at the initial plane. Besides, compared to the Airy beam, the compensating beam owns more caustics distribution at the initial plane and higher energy concentration for its main lobe modulated by the masks. At last, we also showed that the main-lobe orientation of the compensating beam can be adjusted in four different quadrants by altering the signs of the 2D apodization masks. With the novel propagation property and the versatile main-lobe orientation, we believe that the proposed accelerating beam can get special application in the particle manipulation or plasma etc. region.

**Acknowledges**

This work was supported by National Natural Science Foundation of China (11374292 and 1130222).